\documentclass[aps,prl,twocolumn,tightenlines,superscriptaddress]{revtex4}
\usepackage{epsfig,rotating}
\usepackage{multirow}

\newcommand{\nuc}[2]{\hbox{$^{#1}$#2}}

\begin{document}
\title{The structure of \nuc{70}{Fe}: Single-particle and collective degrees of
  freedom}

\author{A.\ Gade}
   \affiliation{National Superconducting Cyclotron Laboratory,
      Michigan State University, East Lansing, Michigan 48824, USA}
   \affiliation{Department of Physics and Astronomy,
      Michigan State University, East Lansing, Michigan 48824, USA}
\author{R.\ V.\ F.\ Janssens}
    \affiliation{Department of Physics and Astronomy, University of North
      Carolina at Chapel Hill, Chapel Hill, North Carolina 27599, USA and
      Triangle Universities Nuclear Laboratory, Duke University, 
      Durham, North Carolina 27708, USA}  
\author{J.\ A.\ Tostevin}
     \affiliation{Department of Physics, University of Surrey,
       Guildford, Surrey GU2 7XH, United Kingdom}
\author{D.\ Bazin}
    \affiliation{National Superconducting Cyclotron Laboratory,
      Michigan State University, East Lansing, Michigan 48824, USA}
      \affiliation{Department of Physics and Astronomy,
      Michigan State University, East Lansing, Michigan 48824, USA}
\author{J.\ Belarge}
\altaffiliation{J. Belarge is currently an MIT Lincoln Laboratory employee. No
  Laboratory funding or resources were used to produce the result/findings
  reported in this publication.} 
    \affiliation{National Superconducting Cyclotron Laboratory,
      Michigan State University, East Lansing, Michigan 48824, USA}
\author{P.\ C.\ Bender}
   \altaffiliation{Present address: Department of Physics, University of Massachusetts Lowell, Lowell, Massachusetts 01854, USA}
 \affiliation{National Superconducting Cyclotron Laboratory,
      Michigan State University, East Lansing, Michigan 48824, USA}
\author{S.\ Bottoni}
\altaffiliation{Present address: Dipartimento di Fisica, Universit\`{a} degli
  Studi di Milano, 20122 Milano, Italy}
     \affiliation{Physics Division, Argonne National Laboratory,
       Argonne, Illinois 60439, USA}
\author{M.\ P.\ Carpenter}
     \affiliation{Physics Division, Argonne National Laboratory,
       Argonne, Illinois 60439, USA}
\author{B.\ Elman}
    \affiliation{National Superconducting Cyclotron Laboratory,
      Michigan State University, East Lansing, Michigan 48824, USA}
    \affiliation{Department of Physics and Astronomy,
      Michigan State University, East Lansing, Michigan 48824, USA}
\author{S.\ J. Freeman}
    \affiliation{School of Physics and Astronomy, University of Manchester,
      Manchester M13 9PL, United Kingdom}
\author{T.\ Lauritsen}
       \affiliation{Physics Division, Argonne National Laboratory,
         Argonne, Illinois 60439, USA}
 \author{S.\ M.\ Lenzi}
        \affiliation{Dipartimento di Fisica e Astronomia dell'Universit\`{a} and
          INFN, Sezione di Padova, I-35131 Padova, Italy}
\author{B.\ Longfellow}
    \affiliation{National Superconducting Cyclotron Laboratory,
      Michigan State University, East Lansing, Michigan 48824, USA}
    \affiliation{Department of Physics and Astronomy,
      Michigan State University, East Lansing, Michigan 48824, USA}
\author{E.\ Lunderberg}
    \affiliation{National Superconducting Cyclotron Laboratory,
      Michigan State University, East Lansing, Michigan 48824, USA}
    \affiliation{Department of Physics and Astronomy,
      Michigan State University, East Lansing, Michigan 48824, USA}
\author{A.\ Poves}
      \affiliation{Departamento de F\'{i}sica Te\'{o}rica e IFT-UAM/CSIC,
        Universidad Aut\'{o}noma de Madrid, E-28049 Madrid, Spain}
\author{L.\ A.\ Riley}
      \affiliation{Department of Physics and Astronomy, Ursinus College,
        Collegeville, Pennsylvania 19426, USA}
\author{D.\ K.\ Sharp}
    \affiliation{School of Physics and Astronomy, University of Manchester,
      Manchester M13 9PL, United Kingdom}
\author{D.\ Weisshaar}
    \affiliation{National Superconducting Cyclotron Laboratory,
      Michigan State University, East Lansing, Michigan 48824, USA}
\author{S.\ Zhu}
     \affiliation{Physics Division, Argonne National Laboratory,
       Argonne, Illinois 60439, USA}
\date{\today}

\begin{abstract}
Excited states in the neutron-rich \nuc{70}{Fe} nucleus were populated
in a one-proton removal reaction from \nuc{71}{Co} projectiles at
87~MeV/nucleon. A new transition was observed with the
$\gamma$-ray tracking array GRETINA and shown to feed the previously
assigned $4^+_1$ state. In comparison to reaction theory calculations
with shell-model spectroscopic factors, it is argued that the new
$\gamma$ ray possibly originates from the $6^+_1$ state. It is further shown
that the Doppler-reconstructed $\gamma$-ray spectra are sensitive
to the very different lifetimes of the $2^+$ and $4^+$  states,
enabling their approximate measurement. The emerging structure of
\nuc{70}{Fe} is discussed in comparison to LNPS-new  large-scale shell-model
calculations.
\end{abstract}

\maketitle

A goal of nuclear science is to achieve an understanding of nuclei and their
properties 
rooted in the fundamental nucleon-nucleon interactions, while
demonstrating predictive 
power for the shortest-lived species located at the fringes of the nuclear
chart. In 
the quest to extrapolate knowledge to the most neutron-rich systems, including
those that may
remain 
beyond experimental reach, much can be learned from nuclei with large neutron
excess that clearly display the effects of structural evolution away from the
valley of stability. Observables measured for such nuclei provide important
extrapolation points toward unknown regions and their successful modeling
offers critical benchmarks for theory. Specifically, the complex interplay
between single-particle and collective degrees of freedom in the nuclear
many-body 
system provides unique and interesting experimental and theoretical challenges.

Neutron-rich \nuc{70}{Fe} is such a nucleus where single-particle structure
is impacted by shell evolution, driven by the spin-isospin parts of the
nucleon-nucleon force, and where significant quadrupole collectivity develops.
In fact, \nuc{70}{Fe} is said to be part of the $N=40$ island of inversion
\cite{Len10,Now16} where the neutron-rich Fe and Cr nuclei around $N=40$
become the most deformed in the region. This is thought to be caused by the
strong quadrupole-quadrupole interaction producing a shape transition in
which highly correlated, many-particle-many-hole configurations become
more bound than the normal-order (spherical) ones~\cite{Len10}. Such islands of
inversion are characterized by rapid structural changes and shape coexistence
\cite{Now16,Gad16a}, providing insight into nuclear structure physics far from stability~\cite{Brown10}. \nuc{70}{Fe} has 12 neutrons more than the heaviest stable
iron isotope, while the heaviest one discovered to date is
\nuc{76}{Fe}~\cite{Sum17}, a nucleus predicted to display collectivity and shape
coexistence~\cite{Now16}  just two protons below \nuc{78}{Ni}. Indeed, within
the iron isotopic chain, \nuc{70}{Fe} is located on the path between the $N=40$
and $N=50$ islands of inversion~\cite{Now16}, with the latter remaining a
challenge for next-generation rare-isotope facilities presently under
construction. \nuc{70}{Fe} has also been used as a seed nucleus in r-process
calculations 
and associated sensitivity studies~\cite{Ani14,Brett12}.  
Spectroscopic information on \nuc{70}{Fe}, limited
to the identification of two states, the first $2^+$ level
and another with a tentative $4^+$ assignment, comes thus far from the
population of excited states in $\beta$ decay~\cite{Ben15}
and a $(p,2p)$ reaction study~\cite{San15}. 

Here, we present the high-resolution spectroscopy of \nuc{70}{Fe}
in the direct one-proton removal reaction \nuc{9}{Be}(\nuc{71}{Co},\nuc{70}{Fe}$+
\gamma$)X at 87~MeV/u, leading to a newly observed $\gamma$-ray transition
and the determination of partial cross sections.
The latter are discussed quantitatively in comparison to
eikonal reaction theory~\cite{knock} with LNPS-new shell-model spectroscopic
factors~\cite{Len10,lnpsU}. The rather simple \nuc{70}{Fe} $\gamma$-ray spectrum observed, with
only 
three peaks, is at odds with the predicted strong population of highly-excited
states. On the experimental side, we propose, as a solution to this puzzle, 
the so-called pandemonium effect~\cite{Har77} arising from a sizable fragmentation
of the  proton
spectroscopic 
strength in \nuc{70}{Fe}. This fragmentation is larger than predicted within the
limited configuration spaces of shell-model calculations, on the theoretical
side. 
While such challenges may actually be rather universal for $\gamma$-ray tagged
direct reactions leading to collective even-even nuclei, it is argued that
observables, such as yrast excitation energies and transition
strengths, are nevertheless well described. From the present data, approximate
lifetimes for the $2^+_1$ and $(4^+_1)$ states were extracted through a Doppler-shift
analysis, and found to be consistent with the results of LNPS-new shell-model
calculations.

The \nuc{71}{Co} secondary beam was produced from projectile fragmentation
of a 140 MeV/u stable \nuc{82}{Se} beam provided by the Coupled Cyclotron
Facility at NSCL~\cite{Gad16b}, impinging on a 446~mg/cm$^2$ \nuc{9}{Be}
production target and separated using a 240~mg/cm$^2$ Al degrader in the A1900
fragment separator~\cite{a1900}. The momentum acceptance of the separator was
restricted to 2\%, yielding on-target rates of typically 65 \nuc{71}{Co}/s.
About 9.5\% of the beam was \nuc{71}{Co}, with \nuc{72,73}{Ni} and
\nuc{74}{Cu} as the most intense components.

The secondary \nuc{9}{Be} reaction target (376~mg/cm$^2$ thick) was located at
the target position of the S800 spectrograph. Reaction products were identified
on an event-by-event basis at the S800 focal plane with the standard detector
systems~\cite{s800}.  The particle identification was performed with the
measured energy loss and time-of-flight information, as 
demonstrated in \cite{Gad14}, for the equivalent reaction on a \nuc{61}{V}
projectile beam. The inclusive cross section for the one-proton removal from
\nuc{71}{Co} to \nuc{70}{Fe} was deduced to be $\sigma_{inc}=11.0(8)$~mb.

\begin{figure}[h]
\epsfxsize 8.2cm
\epsfbox{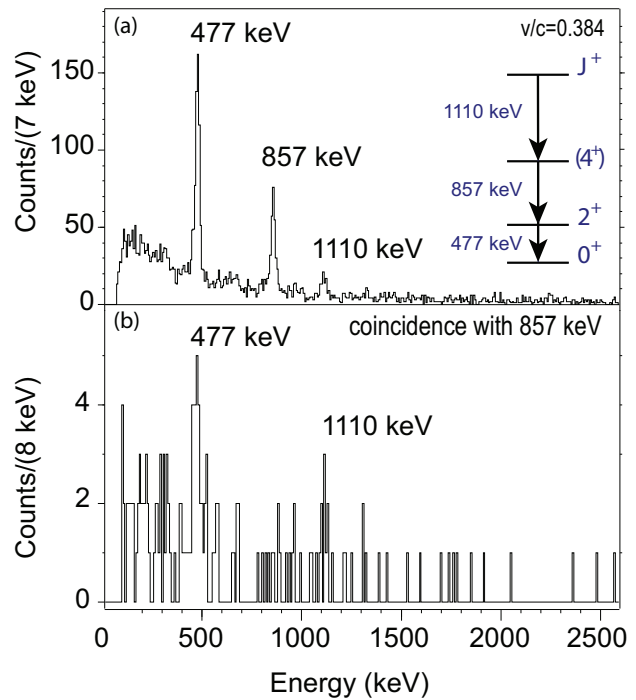}
\caption{\label{fig:gamma} (a) Doppler-reconstructed $\gamma$-ray spectrum in
coincidence with \nuc{70}{Fe} reaction residues. (b)
$\gamma \gamma$ coincidence spectrum gated on the 857-keV transition, returning
peaks at 477 and 1110~keV, leading to the proposed level scheme in the inset of
(a).} 
\end{figure}

The high-resolution $\gamma$-ray detection system GRETINA~\cite{gretina,Wei17},
an array of 36-fold segmented high-purity germanium detectors grouped into
modules of four crystals each, was used to
measure the prompt $\gamma$ rays emitted by the reaction residues. The nine detector modules
available at the time were arranged
in two rings, with four located at 58$^{\circ}$ and five at 90$^{\circ}$
with respect to the beam axis. Online signal decomposition provided $\gamma$-ray
interaction points for event-by-event Doppler reconstruction of the photons
emitted in-flight~\cite{Wei17} at $v/c= 0.4$. The information on the momentum vector of
projectile-like reaction residues, as ray-traced through the spectrograph, was
incorporated into the Doppler reconstruction. Fig.~\ref{fig:gamma}(a) provides
the Doppler-reconstructed $\gamma$-ray spectrum for \nuc{70}{Fe} produced with
nearest-neighbor 
addback included~\cite{Wei17}. The remarkable peak-to-background ratio enabled
 spectroscopy at modest levels of statistics in a nucleus far removed from
 stability. In addition 
to the previously reported $2^+_1 \rightarrow 0^+_1$ and $(4^+_1) \rightarrow
2^+_1$ transitions~\cite{Ben15,San15}, one additional $\gamma$ ray, at
1110(4)~keV, could 
be identified in \nuc{70}{Fe}.

GRETINA's $\gamma\gamma$ coincidence capability resulted in the level
scheme displayed in the inset to Fig.~\ref{fig:gamma}(a).
Figure~\ref{fig:gamma}(b) presents the spectrum in coincidence with the 857-keV
transition, returning the other two $\gamma$ rays. Together with the peak
intensities, this places the three transitions in a cascade as indicated in
Fig.~\ref{fig:gamma}(a).

The photopeak efficiency of GRETINA was calibrated with standard sources and
corrected for the Lorentz boost of the $\gamma$-ray distribution emitted by the
residual nuclei moving at almost 40\% of the speed of light. The fact that one
crystal in a forward detector module was not working was taken into account. The
peak 
areas were determined from the spectrum of \nuc{70}{Fe} without addback, avoiding
uncertainties associated with the addback efficiency~\cite{Wei17}.
Partial proton-removal cross sections to the specific final states were determined
from the efficiency-corrected $\gamma$-ray peak areas, with discrete feeding
subtracted, relative to the number of incoming \nuc{71}{Co} projectiles and
the number density of the target: $\sigma(0^+)=1.0(6)$~mb, $\sigma(2^+)=4.0(8)
$~mb, $\sigma(4^+)=4.1(6)$~mb, and $\sigma(J^+)=1.85(30)$~mb.

One-nucleon removal is a direct reaction with sensitivity to single-particle
degrees of freedom. The cross sections for the population of individual \nuc{70}{Fe}
final states depend sensitively on the projectile to final-state one-body overlaps
and on their normalizations; i.e., the spectroscopic factors~\cite{knock}. Shell-model
calculations with the LNPS-new effective interaction predict a $7/2^-$ \nuc{71}{Co}
ground state, in agreement with $\beta$-decay results~\cite{Raj12}, and a
low-lying (200~keV) $1/2^-$
isomer that has not yet been observed.

\begin{figure}[h]
\epsfxsize 8.2cm
\epsfbox{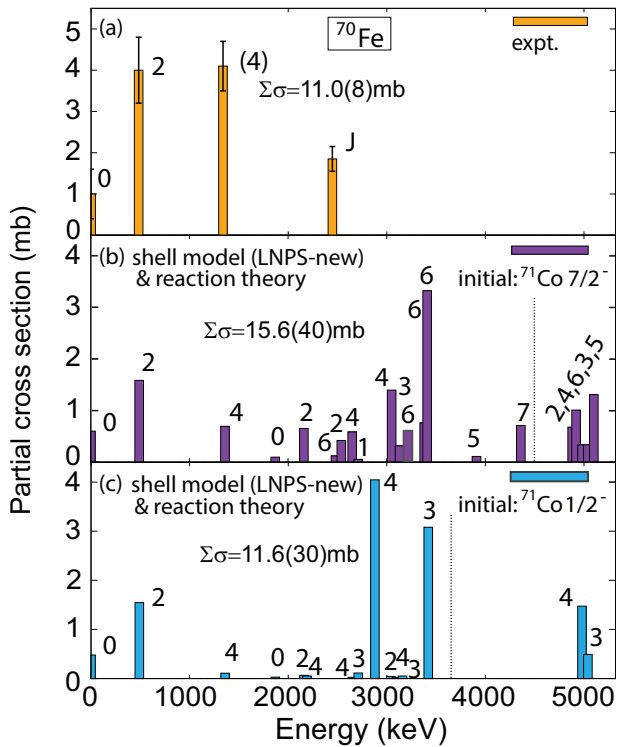}
\caption{\label{fig:xsec} Partial proton removal cross sections from
  \nuc{71}{Co} for the population
of positive-parity \nuc{70}{Fe} final states: experiment (a)
and calculations assuming the $7/2^-$ (b) and $1/2^-$ (c) shell-model
initial states of the \nuc{71}{Co} projectile and $R_s$. The calculated cross sections
indicated beyond the dashed lines correspond to the summed strengths to bound
states for the given $J^+$ values at high excitation energy, placed here at 5~MeV. }
\end{figure}

Using the one-nucleon removal methodology detailed in Ref.~\cite{Gad08} together
with the LNPS-new~\cite{Len10,lnpsU} spectroscopic factors for incident \nuc{71}{Co} in
the $7/2^-$ and $1/2^-$ states, the partial cross sections to bound \nuc{70}{Fe}
shell-model final states were calculated and confronted with experiment in Fig.~\ref{fig:xsec}. With reference to the nucleon removal reaction systematics 
\cite{Tos14}, a reduction factor $R_s = 0.4(1)$ was assumed between the calculated 
and the measured cross sections, based on the final-states yields-weighted proton 
separation energy, $S_p \approx 18$~MeV, resulting in a proton and neutron
separation  
energy asymmetry of $\Delta S=S_p -  S_n \approx$~12 MeV for \nuc{71}{Co} 
\cite{AME2012}. The presence of both the ground and the isomeric state in the 
incoming \nuc{71}{Co} beam cannot be ruled out and the measured cross section 
distribution may correspond to a linear combination of both.

For both possible initial states, the predicted population pattern is at odds with that
measured and with the simple $\gamma$-ray spectrum observed. A strong population
of high-lying states, such as the $6^+_4$, $4^+_4$ or $3^+_3$ levels, would lead
to the presence of several strong additional transitions, connecting the populated high-lying states to the
level scheme reported here. For each assumed \nuc{71}{Co} initial state, the sums
of the partial cross sections to all bound shell-model final states below $S_n=
5.32(64)$~MeV~\cite{AME2012} are $\sigma_{inc}^{7/2-}=15.6(40)$~mb and $\sigma_{inc}
^{1/2-}=11.6(30)$~mb, slightly higher than the measured inclusive cross section
of $\sigma_{inc}=11.0(8)$~mb.

The apparent simplicity of the observed population of final states in
\nuc{70}{Fe} is rather puzzling. We note that the $\gamma$-ray spectrum reported here is consistent with
that reported from the $(p,2p)$ reaction, where no $\gamma$ rays other
than those associated with the $2^+_1 \rightarrow 0^+_1$ and $(4^+_1)
\rightarrow 2^+_1$ transitions were
observed~\cite{San15}. While the cross sections from our \nuc{9}{Be}-induced
proton removal and $(p,2p)$ may differ quantitatively, qualitatively they will populate
the same  
proton-hole configurations and the respective cross sections should scale with the
same 
spectroscopic factors. The modest energy resolution accomplished with a scintillator array in
the $(p,2p)$ measurement of Ref.~\cite{San15} likely prevented the identification of the
(weak) 1110-keV peak due to a disadvantage in the peak-to-background ratio.
However, their superior detection efficiency should have enabled the clear observation
of intense feeding transitions from high-lying states in view of their predicted
strong population. Such concentration of proton spectroscopic strength
in low-lying yrast states in the $N=40$ region has also been reported for other
proton removal 
reactions leading to \nuc{66,68}{Fe}~\cite{Adr08}, \nuc{60}{Ti}~\cite{Gad14} and
\nuc{66}{Cr}, \nuc{72}{Fe}~\cite{San15}.

One must consider the pandemonium effect~\cite{Har77}, a situation where
modestly 
efficient $\gamma$-ray spectroscopy of discrete transitions misses the population
of high-lying states ultimately de-exciting to the yrast states through a large number
of weak transitions. This effect, thus, attributes high-lying strengths
to the yrast states that act as collectors for weak feeding transitions escaping
observation. This is a possibility given the extreme level density predicted
by the shell-model - with more than 100 states below $S_n=5.32$~MeV in \nuc{70}{Fe} --
but would actually require a larger fragmentation of the strength than that
predicted. Specifically, further fragmentation would be expected beyond that to one or two high-lying
states as the latter would certainly have their strongest transitions observed. Such a
scenario of sizable fragmentation could also explain the slight mismatch between the
calculated and measured inclusive cross sections, as increased fragmentation would likely
shift spectroscopic strength to energies beyond $S_n$. Hence, an understanding
of spectroscopic
strengths in even-even nuclei of the $N=40$ island of inversion may demand yet larger model
spaces and more complex, mixed configurations, requiring the inclusion of
orbitals beyond $\nu
g_{9/2}$ and $\nu d_{5/2}$ that were already identified as critical for
describing the region~\cite{Len10}.
Assuming such an interpretation of the present data, we suggest the newly established
level at 2448(4)~keV to correspond to the $(6^+_1)$ state or a higher-lying $4^+$ state. The energies of the transitions from the
$2^+$ and $4^+$ states reported in the $\beta$-decay work~\cite{Ben15} were used
here to deduce the level energy due to a significant sensitivity to excited-state
lifetimes in the present in-beam data, as discussed below. For the two possible \nuc{71}{Co}
initial states, the strongly populated $6^+$ levels ($7/2^-$ initial state) and $4^+$
and $3^+$ levels ($1/2^-$ isomeric initial state) would, in a pandemonium picture,
ultimately feed into the yrast $6^+_1$ and $4^+$ states. We note the
very good agreement with the LNPS-new shell-model calculation that places the $6^+_1$
state at 2.48~MeV, within 30~keV of the measured value proposed here, while the
closest higher-excited $4^+$ is predicted to be located 200~keV higher.  

\begin{figure}[h]
\epsfxsize 8.4cm
\epsfbox{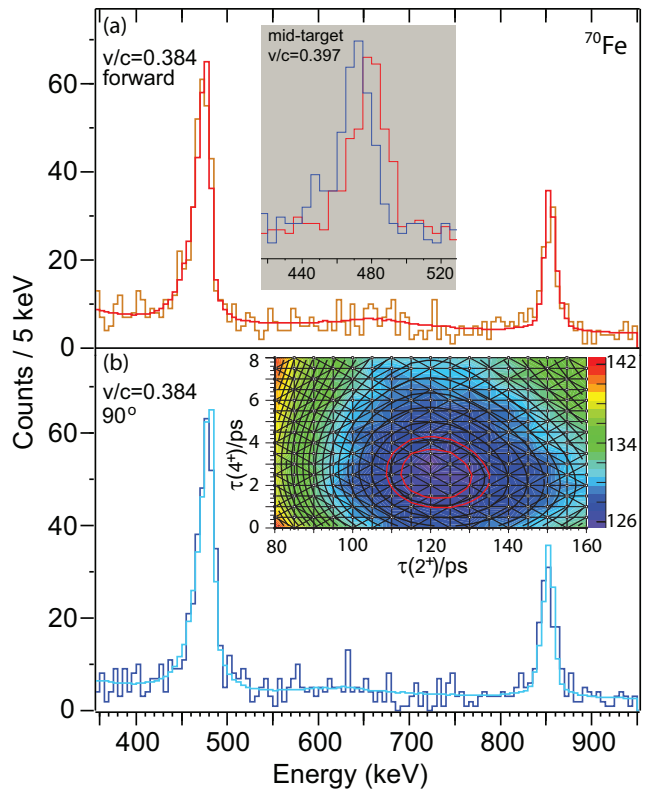}
\caption{\label{fig:LT} Doppler-corrected $\gamma$-ray spectra from GRETINA's forward
($58^{\circ}$) and $90^{\circ}$ rings ($v/c=0.384$; the $(4^+_1) \rightarrow
  2^+_1$ transition lines up in both rings). A significant energy difference is
  observed 
for the $2^+_1 \rightarrow 0^+_1$ transition at the mid-target $v/c$ (top
inset). Overlaid  
are GEANT simulations that minimize a negative log-likelihood surface (bottom inset) 
of a fit to a large set of simulated lifetimes properly accounting for the $4^+_1$ 
feeding of the $2^+_1$ state.}
\end{figure}

\begin{figure}[h]
\epsfxsize 8.4cm
\epsfbox{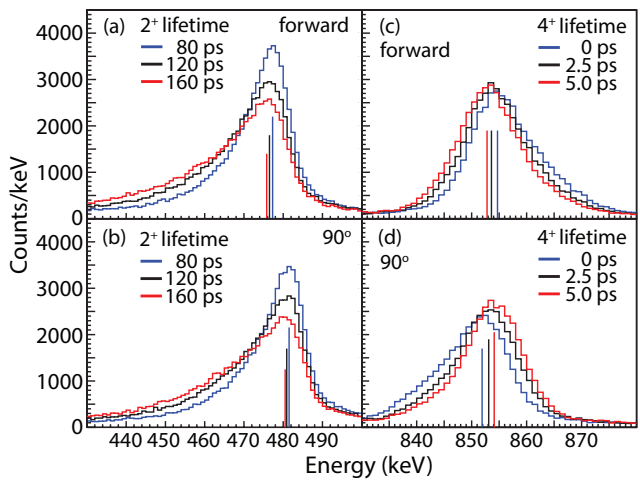}
\caption{\label{fig:LTsim} Line shapes simulated with GEANT for different
  lifetimes of the $2^+_1 \rightarrow 0^+_1$ transition in the  (a) forward and
  (b) $90^{\circ}$ detectors and for various lifetimes of the $(4^+_1) \rightarrow
  2^+_1$ transition in the  (c) forward and (d) $90^{\circ}$ detectors ($v/c=0.384$). This
  illustrates the specific sensitivities that the present measurement exhibits
  for the long $\tau(2^+)$ (strong tails) and short $\tau(4^+)$ (shifting
  peak position - indicated by vertical lines) values.   }
\end{figure}

Unlike any other shell-model effective interaction for this mass region, LNPS(-new)
\cite{Len10,lnpsU} has demonstrated predictive power for collective observables,
such as for the $B(E2)$ transition strengths and energies of the lowest-lying
$2^+$ 
and $4^+$ states~\cite{Len10,Gad10,Rot11,Gad14,San15}. For the measurements
reported here, the $\gamma$-ray spectra reveal effects 
attributed to excited-state lifetimes that
can inform on the expected collectivity of \nuc{70}{Fe}. The inset of Fig.~\ref{fig:LT}(a) demonstrates a distinct shift in energy of the $2^+_1 \rightarrow 0^+_1$
transition detected in the GRETINA detectors mounted in the $58^{\circ}$ and
$90^{\circ}$ rings when corrected for the Doppler shift assuming the mid-target
beam velocity. This  indicates that the $2^+_1$ state decays on average with a
velocity lower than the mid-target one; i.e., when the \nuc{70}{Fe} ions lost
more energy. 
This is also the reason for the mismatch between the transition $\gamma$-ray
energy reported here and that from $\beta$ decay, 477~keV
vs. 483~keV. Using GEANT~\cite{geant}, the lifetimes can be determined by
matching to simulations the peak shapes and peak positions observed in detector
groups at different polar angles, such as forward and $90^{\circ}$. We note that
for long lifetimes the peak shape and peak position are impacted while shorter
lifetimes largely affect the peak position only. Essential for this simulation
is the precise knowledge of the transition energy and the target position along
the beam axis. A 
target offset of 0.3(3)~mm downstream from the center of GRETINA was determined
with the help of a known $\gamma$-ray transition in a contaminant. This value
is small as compared to the actual $\approx 2$~mm target thickness. Using
the transition  
energies of 483 and 855~keV from $\beta$ decay~\cite{Ben15} and 
the target offset, effective lifetimes for the $2^+_1$ and $4^+_1$ states were
extracted from a log likelihood minimization procedure that takes into account
the feeding by the $4^+$ level (the 1110~keV transition was too weak for such an 
analysis and was assumed to be prompt). The results are shown in
Fig.~\ref{fig:LT},  
where the spectra for each ring of GRETINA are overlaid with the GEANT simulation 
that minimized the negative log-likelihood surface, given as an inset
(Fig.~\ref{fig:LT}(b)). To illustrate the sensitivity of the present measurement
to the different lifetimes in more detail, Fig.~\ref{fig:LTsim} provides
simulated line shapes for the two transitions of interest for various lifetime
values. For longer lifetimes, the primary sensitivity is to the tails of the peak shape, while
shorter lifetimes affect the positions of the peak maximum. We note that similar
sensitivity 
studies for other beam and target combinations simulated for the AGATA array can be found in~\cite{Dom12}.
      
The deduced effective lifetimes (95\% confidence interval for the fit) are $\tau_{
\mathrm{eff}}(2^+_1)=120^{+15}_{-11}$~ps and $\tau_{\mathrm{eff}}(4^+_1)=2.3 \pm
1.5$~ps, respectively. Adding the systematic uncertainty from the target offset increases the
error range of the longer lifetime to $\tau_{\mathrm{eff}}(2^+_1)=120 \pm 20$~ps.
These lifetimes have to be considered as effective ones since the yrast
cascade is, most likely, subject to significant unobserved feeding from higher-lying
states which could lead to an overestimation of the lifetimes. Given
these uncertainties, one may conservatively conclude that the observed lifetime
effects are consistent with a $\tau(2^+_1)$ value of order 100~ps and $\tau(4^+_1)
\approx 2 - 4$~ps. This is in broad agreement with the LNPS shell-model calculations
of the corresponding $B(E2)$ values quoted in Ref.~\cite{San15} from which $\tau(2^+_1)=81$~ps
and $\tau(4^+_1)=3$~ps are extracted when using the measured transition
energies. This underlines the collective nature of \nuc{70}{Fe} as well as the
success of the LNPS shell-model calculations~\cite{Len10,lnpsU} in the
description of this hallmark property of a nucleus within the island of
inversion. 

In summary,  high-resolution in-beam $\gamma$-ray spectroscopy with GRETINA was performed
for the neutron-rich nucleus \nuc{70}{Fe} following  one-proton removal from \nuc{71}{Co}
projectiles. A newly observed 1101(4)-keV $\gamma$ ray was tentatively assigned
to the transition from the $(6^+_1)$ or a higher-lying $(4^+)$ state at 2.448(4)~MeV to the $(4^+_1)$ level,
based on comparison with the results of calculations using eikonal reaction
theory incorporating spectroscopic factors from shell-model calculations based
on the LNPS-new effective interaction. The $(J^+) \rightarrow (4^+_1) \rightarrow 2^+_1$ cascade is found to agree well with the shell-model description if the newly discovered level is the $6^+_1$ state. The
pandemonium effect 
and an implied large fragmentation of spectroscopic strength are proposed to account
for the marked discrepancy between measured and calculated population patterns:
these present a challenge from an experimental and theoretical point of
view. Despite these limitations, it was shown that besides the excitation
energies, the shell-model calculations also account for another collective
observable, the approximate excited-state lifetimes of the $2^+_1$ and $(4^+_1)$
states, extracted here via Doppler shifts and line shapes.

\begin{acknowledgments}
This work was supported by the US National Science Foundation (NSF) under Cooperative
Agreement No.  PHY-1565546 (NSCL) and grant No. PHY-1617250 (Ursinus), by the US Department of Energy (DOE) National Nuclear Security
Administration under award numbers DE-NA0003180 and DE-NA0000979, and by the DOE-SC Office
of Nuclear Physics under Grants DE-FG02-08ER41556 (NSCL), DE-FG02-97ER41041
(UNC), DE-FG02-97ER41033 (TUNL), and DE-AC02-06CH11357 (ANL). GRETINA was funded by the DOE,
Office of Science. Operation of the array at NSCL was supported by the DOE under Grant No.
DE-SC0014537 (NSCL) and DE-AC02-05CH11231 (LBNL). J.A.T. and S.J.F \& D.K.S
acknowledge the support 
of the Science and Technology Facilities Council (UK) grant ST/L005743/1 and
ST/L005794/1, respectively. We also thank T.J. Carroll for the use of the Ursinus College Parallel Computing Cluster,
supported by NSF grant No. PHY-1607335. A.P. is supported in part by MINECO
(Spain) Grant FPA2014-57196 and the Severo Ochoa Programme SEV-2016-0597.

\end{acknowledgments}


\begin{thebibliography}{10}

\bibitem{Len10} S.\ M.\ Lenzi, F.\ Nowacki, A.\ Poves, and K.\ Sieja
Phys.\ Rev.\ C 82, 054301 (2010).

\bibitem{Now16} F. Nowacki, A. Poves, E. Caurier, B. Bounthong,
Phys. Rev. Lett. 117, 272501 (2016).

\bibitem{Gad16a} A. Gade and S. N. Liddick, J. Phys. G 43,
  024001 (2016).


\bibitem{Brown10} B.\ Alex Brown, Physics 3, 104 (2010).



\bibitem{Sum17} T. Sumikama, S. Nishimura, H. Baba, F. Browne, P. Doornenbal,
  N. Fukuda, S. Franchoo, G. Gey, N. Inabe, T. Isobe, P.R. John, H.S. Jung, D. Kameda,
  T. Kubo, Z. Li, G. Lorusso, I. Matea, K. Matsui, P. Morfouace, D. Mengoni,
  D.R. Napoli, M. Niikura, H. Nishibata, A. Odahara, E. Sahin, H. Sakurai,
  P.-A. Soderstrom, G.I. Stefan, D. Suzuki, H. Suzuki, H. Takeda, R. Taniuchi,
  J. Taprogge, Zs. Vajta, H. Watanabe, V. Werner, J. Wu, Z.Y. Xu, A. Yagi, K. Yoshinaga,  Phys. Rev. C 95, 051601 (2017)

\bibitem{Ani14} A. Aprahamian, I. Bentley, M. Mumpower, and R. Surman, AIP ADVANCES 4, 041101 (2014).

\bibitem{Brett12} S. Brett, I. Bentley, N. Paul, R. Surman, and A. Aprahamian, Eur. Phys. J. A 48, 184 (2012).

\bibitem{Ben15} G. Benzoni, A.I. Morales, H. Watanabe, S. Nishimura, L. Coraggio,
  N. Itaco, A. Gargano, F. Browne, R. Daido, P. Doornenbal, Y. Fang, G. Lorusso,
  Z. Patel, S. Rice, L. Sinclair, P.-A. Soderstrom, T. Sumikama, J. Wu, Z.Y. Xu,
  R. Yokoyama, H. Baba, R. Avigo, F.L. Bello Garrote, N. Blasi, A. Bracco, F. Camera,
  S. Ceruti, F.C.L. Crespi, G. de Angelis, M.-C. Delattre, Zs. Dombradi, A. Gottardo,
  T. Isobe, I. Kuti, K. Matsui, B. Melon, D. Mengoni, T. Miyazaki,
  V. Modamio-Hoybjor, Phys. Lett. B 751, 107 (2015).

\bibitem{San15} C. Santamaria, C. Louchart, A. Obertelli, V. Werner,
  P. Doornenbal, F. Nowacki, G. Authelet, H. Baba, D. Calvet, F. Chateau,
  A. Corsi, A. Delbart, J.-M. Gheller, A. Gillibert, T. Isobe, V. Lapoux,
  M. Matsushita, S. Momiyama, T. Motobayashi, M. Niikura, H. Otsu, C. Peron,
  A. Peyaud, E. C. Pollacco, J.-Y. Rousse, H. Sakurai, M. Sasano, Y. Shiga,
  S. Takeuchi, R. Taniuchi, T. Uesaka, H. Wang, K. Yoneda, F. Browne,
  L. X. Chung, Zs. Dombradi, S. Franchoo, F. Giacoppo, A. Gottardo,
  K. Hadynska-Klek, Z. Korkulu, S. Koyama, Y. Kubota, J. Lee, M. Lettmann,
  R. Lozeva, K. Matsui, T. Miyazaki, S. Nishimura, L. Olivier, S. Ota, Z. Patel,
  N. Pietralla, E. Sahin, C. Shand, P.-A. S\"oderstr\"om, I. Stefan,
  D. Steppenbeck, T. Sumikama, D. Suzuki, Zs. Vajta, J. Wu, and Z. Xu,
  Phys. Rev. Lett. 115, 192501 (2017).

\bibitem{knock}   J.\ A.\ Tostevin, J.\ Phys.\ G: Nucl.\ Part.\ Phys.\ 25, 735 (1999)
  and P.\ G.\ Hansen and J.\ A.\ Tostevin, Annu. Rev. Nucl. Part. Sci. 53, 221
  (2003).

\bibitem{lnpsU} F. Nowacki (private communication).

\bibitem{Har77} J.\ C.\ Hardy, L.\ C. Carraz, B.\ Jonson, P.\ G.\ Hansen, Phys.\
  Lett.\ B 71, 307 (1977).

\bibitem{Gad16b} A. Gade and B. M. Sherrill, Phys. Scr. 91, 053003 (2016).

\bibitem{a1900} D.\ J.\ Morrissey {\it et al.}, Nucl.\ Instrum.\ Methods
  in Phys.\ Res.\ B {\bf 204}, 90 (2003).

\bibitem{s800} D.\ Bazin {\it et al.}, Nucl.\ Instrum.\ Methods in Phys.\
  Res.\ B {\bf 204}, 629 (2003).

\bibitem{Gad14} A. Gade, R. V. F. Janssens, D. Weisshaar, B. A. Brown,
  E. Lunderberg, M. Albers, V. M. Bader, T. Baugher, D. Bazin, J. S. Berryman,
  C. M. Campbell, M. P. Carpenter, C. J. Chiara, H. L. Crawford, M. Cromaz,
  U. Garg, C. R. Hoffman, F. G. Kondev, C. Langer, T. Lauritsen, I. Y. Lee,
  S. M. Lenzi, J. T. Matta, F. Nowacki, F. Recchia, K. Sieja, S. R. Stroberg,
  J. A. Tostevin, S. J. Williams, K. Wimmer, and S. Zhu, Phys. Rev. Lett. 112,
  112503 (2014).

\bibitem{gretina} S.\ Paschalis et al., Nucl.\ Instrum.\ Methods Phys.\ Res.\ A
709, 44 (2013).

\bibitem{Wei17} D. Weisshaar, D. Bazin, P. C. Bender, C. M. Campbell,
  F. Recchia, V. Bader, T. Baugher, J. Belarge, M. P. Carpenter, H. L. Crawford,
  M. Cromaz, B. Elman, P. Fallon, A. Forney, A. Gade, J. Harker, N. Kobayashi,
  C. Langer, T. Lauritsen, I. Y. Lee, A. Lemasson, B. Longfellow, E. Lunderberg,
  A. O. Macchiavelli, K. Miki, S. Momiyama, S. Noji, D. C. Radford, M. Scott,
  J. Sethi, S. R. Stroberg, C. Sullivan, R. Titus, A. Wiens, S. Williams,
  K. Wimmer, and S. Zhu, Nuclear Instrum. Methods Phys. Res. A 847, 187 (2017).

\bibitem{Raj12} M. M. Rajabali, R. Grzywacz, S. N. Liddick, C. Mazzocchi,
  J. C. Batchelder, T. Baumann, C. R. Bingham, I. G. Darby, T. N. Ginter,
  S. V. Ilyushkin, M. Karny, W. Krolas, P. F. Mantica, K. Miernik, M. Pf\"utzner,
  K. P. Rykaczewski, D. Weisshaar, and J. A. Winger, Phys. Rev. C 85, 034326 (2012)

\bibitem{Gad08} A. Gade, P. Adrich, D. Bazin, M. D. Bowen, B. A. Brown,
  C. M. Campbell, J. M. Cook, T. Glasmacher, P. G. Hansen, K. Hosier,
  S. McDaniel, D. McGlinchery, A. Obertelli, K. Siwek, L. A. Riley,
  J. A. Tostevin, and D. Weisshaar , Phys. Rev. C 77, 044306 (2008).

\bibitem{Tos14} J. A. Tostevin and A. Gade, Phys. Rev. C 90, 057602 (2014).

\bibitem{AME2012} M. Wang, G. Audi, A.H. Wapstra, F.G. Kondev, M. MacCormick, X. Xu,
  B. Pfeiffer, Chin. Phys. C 36, 1603 (2012).

\bibitem{Adr08} P. Adrich, A. M. Amthor, D. Bazin, M. D. Bowen, B. A. Brown,
  C. M. Campbell, J. M. Cook, A. Gade, D. Galaviz, T. Glasmacher, S. McDaniel,
  D. Miller, A. Obertelli, Y. Shimbara, K. P. Siwek, J. A. Tostevin, and
  D. Weisshaar, Phys. Rev. C 77, 054306 (2008).

\bibitem{Gad10} A. Gade, R. V. F. Janssens, T. Baugher, D. Bazin, B. A. Brown,
  M. P. Carpenter, C. J. Chiara, A. N. Deacon, S. J. Freeman, G. F. Grinyer,
  C. R. Hoffman, B. P. Kay, F. G. Kondev, T. Lauritsen, S. McDaniel,
  K. Meierbachtol, A. Ratkiewicz, S. R. Stroberg, K. A. Walsh, D. Weisshaar,
  R. Winkler, and S. Zhu, Phys. Rev. C 81, 051304(R) (2010).

\bibitem{Rot11} W. Rother, A. Dewald, H. Iwasaki, S. M. Lenzi, K. Starosta,
  D. Bazin, T. Baugher, B. A. Brown, H. L. Crawford, C. Fransen, A. Gade,
  T. N. Ginter, T. Glasmacher, G. F. Grinyer, M. Hackstein, G. Ilie, J. Jolie,
  S. McDaniel, D. Miller, P. Petkov, Th. Pissulla, A. Ratkiewicz, C. A. Ur,
  P. Voss, K. A. Walsh, D. Weisshaar, and K.-O. Zell, Phys. Rev. Lett. 106,
  022502 (2011).

\bibitem{geant} UCGretina GEANT4, L.\ A.\ Riley, Ursinus College, unpublished.

\bibitem{Dom12} C. Domingo-Pardo, D. Bazzacco, P. Doornenbal, E. Farnea,
  A. Gadea, J. Gerl, H. J. Wollersheim, Nucl.\ Instrum.\ Methods Phys.\ Res.\ A
694, 297 (2012).


\end{thebibliography}
\end{document}